\documentclass[useAMS,usenatbib]{mn2e}
\usepackage{psfig}
\usepackage{amssymb}
\usepackage{color}
\usepackage{enumerate}
\usepackage{rotating}
\usepackage{amsmath}
\usepackage{mathrsfs}
\usepackage{ragged2e}
\usepackage{hyperref}
\usepackage{amsmath}

\title[The dependencies of halo bias]
{The dependence of halo bias on age, concentration and spin}

\author[Sato-Polito et al.]{
\parbox[t]{\textwidth}{
Gabriela Sato-Polito$^{1}$\thanks{E-mail: gabriela.satopolito@usp.br}, Antonio D. Montero-Dorta$^{1}$, L. Raul Abramo$^{1}$, Francisco Prada$^{2}$ \& Anatoly Klypin$^{3}$}
\vspace*{6pt} \\
$^1$ Departamento de F\'isica Matem\'atica, Instituto de F\'isica, Universidade de S\~ao Paulo, Rua do Mat\~ao 1371, CEP 05508-090, \\
S\~ao Paulo, Brazil \\
$^2$ Instituto de Astrof{\'i}sica de Andaluc{\'i}a (CSIC), Glorieta de la Astronom{\'i}a, E-18080 Granada, Spain \\
$^3$ Astronomy Department, New Mexico State University, Las Cruces, NM, USA \\
\vspace{-0.4cm}
}

\date{Accepted ---. Received ---;in original form --- \vspace{-0.3cm}}

\def\simlt{\lower.5ex\hbox{$\; \buildrel < \over \sim \;$}}
\def\simgt{\lower.5ex\hbox{$\; \buildrel > \over \sim \;$}}
\usepackage{graphicx}
\usepackage{rotating}

\definecolor{red}{rgb}{1,0,0}

\begin{document}

\bibliographystyle{mnras}

\maketitle

\begin{abstract}

Halo bias is the main link between the matter distribution and dark matter halos. In its simplest form, halo bias is determined by halo mass, but there are known additional dependencies on other halo properties which are of consequence for accurate modeling of galaxy clustering. Here we present the most precise measurement of these {\em secondary-bias} dependencies on halo age, concentration, and spin, for a wide range of halo masses spanning from 10$^{10.7}$ to 10$^{14.7}$ $h^{-1}$ M$_{\odot}$. At the high-mass end, we find no strong evidence of assembly bias for masses above M$_{vir}$ $\sim10^{14}$ $h^{-1}$ M$_{\odot}$. Secondary bias exists, however, for halo concentration and spin, up to cluster-size halos, in agreement with previous findings. For halo spin, we report, for the first time, two different regimes: above M$_{vir}\sim$10$^{11.5}$ $h^{-1}$ M$_{\odot}$, halos with larger values of spin have larger bias, at fixed mass, with the effect reaching almost a factor 2. This trend reverses below this characteristic mass. In addition to these results, we test, for the first time, the performance of a multi-tracer method for the determination of the relative bias between different subsets of halos. We show that this method increases significantly the signal-to-noise of the secondary-bias measurement as compared to a traditional approach. This analysis serves as the basis for follow-up applications of our multi-tracer method to real data.

\end{abstract}


\section{Introduction}
The clustering of galaxies is the prime observable that can be used to trace the large-scale structure of the Universe (LSS). In the standard model of cosmology, dark matter clusters along density peaks that were generated during inflation and collapse to form dark matter halos. In this scenario, galaxies form when gas falls into collapsing dark-matter halos (e.g., \citealt{white1991}). Hence, the relationship between galaxies, halos, and the underlying matter distribution is crucial to our ability to test cosmological and galaxy formation models against observations.

The bias of dark matter halos can be broadly defined as the relation between the distribution of halos and the underlying matter density field. In its simplest description, the linear halo bias depends only on halo mass, with more massive halos being more strongly clustered than less massive halos \citep{Kaiser1984}. However, halo bias is a much more complex effect that is known to depend on a variety of secondary halo properties. Among these dependencies on secondary properties, the most studied is the dependence on assembly history, called {\it{halo assembly bias}}. Low mass halos (M$<$M$_{\text{*}}$, where M$_{\text{*}}$ is the characteristic mass scale) that assemble a significant portion of their mass early on were shown to be more tightly clustered than halos that assemble at later times, {\it{at fixed halo mass}} (see, e.g., \citealt{gao2005, wechsler2006, li2008,salcedo2018,han2018}). 
For high mass halos (M$>$M$_{\text{*}}$), however, the 
picture is less clear, with most studies showing a small or absent assembly bias 
signal at M$\sim 10^{14}$ $h^{-1}$ M$_{\odot}$ \citep{gao2007, salcedo2018}. For cluster-size halos of  
M$\sim 10^{15}$ $h^{-1}$ M$_{\odot}$, \citet{chue2018} report a significant detection of 
halo assembly bias, of the opposite sign to the one observed at small halo masses (i.e., 
halos are more tightly clustered than their older counterparts), with the 
inversion occurring precisely at M$\sim 10^{14}$ $h^{-1}$ $M_{\odot}$. These conclusions 
appear to be in contradiction with the results from \citet{mao2018}, who claimed to have 
found no halo assembly bias for very massive halos. The analysis of halo assembly bias is extended to higher orders in \cite{Angulo2008}. \citet{li2008} and \citet{chue2018} further discussed the dependence of the halo assembly bias signal on the particular definition 
of halo age/formation time.

\begin{table}
   \centering
   \caption{Numerical properties of MultiDark simulations. The columns correspond to the name of the simulation, the length of the box's side, the number of particles, the force resolution, and the mass of each simulated particle.}
   \begin{tabular}{c|c|c|c|c}
      \hline
      Name & $L_{box}$  & $N_p$ & $\epsilon$ & $M_p$  \\
      & (h$^{-1}$ Gpc) & & (h$^{-1}$ kpc) & (h$^{-1}$ M$_{\odot}$) \\ \hline
      SMDPL & 0.4 & 3840$^3$ & 1.5 & 9.6 $\times 10^7$ \\
      MDPL2 & 1 & 3840$^3$ & 5 & 1.5 $\times 10^9$ \\
      BigMDPL & 2.5 & 3840$^3$ & 10 & 2.4 $\times 10^{10}$ \\
      HugeMDPL &  4.0  & 4096$^3$ & 25 & 7.9 $\times 10^{10}$\\ \hline
   \end{tabular}
   \label{table:sims}
\end{table}

\begin{figure*}
  \centering
    \includegraphics[width=0.9\textwidth]{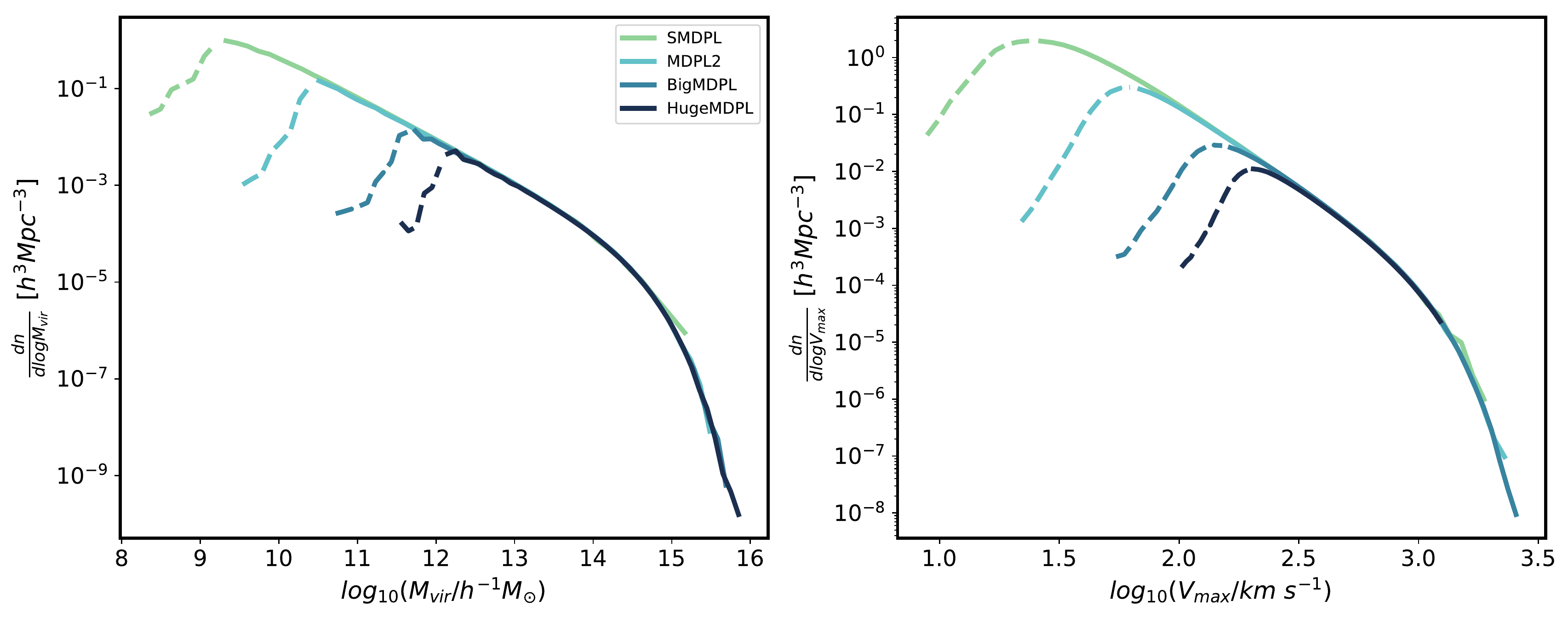}
  \caption{Halo mass function (left) and velocity function (right) for the 4 MultiDark simulation boxes at $z=0$ employed in this work (see Table~\ref{table:sims} for their numerical parameters). This figure illustrates the vast range of halo masses, velocities and abundances analyzed in this work. Dashed lines show the range of halo masses and velocities where completeness drops due to mass resolution effects. For the more conservative resolution limits employed in this work, see Table~\ref{table:mcuts}.}
  \label{fig:mass_function}
\end{figure*}

Halo bias has been shown to depend on a number of other halo properties, including spin, 
concentration, and shape. Hence the term {\it{halo assembly bias}} has progressively been replaced 
by the more general {\it{secondary bias}}\footnote{Throughout the text, the term ``assembly bias'' will exclusively refer to secondary bias produced by age, i.e. formation epoch. The effects produced by concentration and spin will be dubbed spin and concentration bias, respectively.}. Concentration, in particular, has been extensively 
used as a proxy for formation epoch (see \citealt{wechsler2002}). Although the behavior observed for concentration is similar to that reported for age, there are some qualitative differences. At halo masses of M$\sim 10^{13}~h^{-1}M_{\odot}$, a change of regime is well
established: halo bias is larger for more concentrated halos below this mass, but the trend
reverses for higher masses (see, e.g., \citealt{wechsler2006,gao2007,salcedo2018,han2018}). The other ``secondary" property that has drawn significant attention in recent years is spin, $\lambda$, 
which is proportional to the angular momentum of the halo. At fixed mass, halos with larger
values of $\lambda$ are found to be more tightly clustered than those having smaller values 
across the entire mass range considered, which typically covers M$\gtrsim 10^{12}~ h^{-1} M_{\odot}$ (e.g., \citealt{gao2007,Bett2007,Faltenbacher2010,Lacerna2012}). The effect, however, appears to increase at the high-mass end \citep{salcedo2018}.

Despite the variety of measurements, a comprehensive physical model of secondary halo bias is yet to be established. \cite{dalal2008} proposed that certain features of halo assembly bias for high-mass halos can be understood through the statistics of primordial density peaks. At low masses, the authors argue that assembly bias arises from a subpopulation of low-mass halos whose mass accretion has ceased. For other attempts, we refer the reader to \cite{Zentner2007} and \cite{Sendvik2007}, which are based on the implementation of the ellipsoidal collapse model in the framework of the excursion set formalism.

Secondary halo biases have important consequences for the modeling of 
galaxy clustering. Measurements of the two-point correlation function from surveys like 
the Sloan Digital Sky Server \citep{sdss2000} or the 2dF Galaxy Redshift Survey \citep{2dfgrs2001} indicate that more massive, more luminous, and redder galaxies are, in general, 
more tightly clustered than their less massive, less luminous, and bluer counterparts (e.g., \citealt{zehavi2005,Guo2014}). The establishment of secondary bias invalidates any simple 
conclusion exclusively based on halo mass, consequently forcing halo-galaxy connection 
models to adjust (see \citealt{hearin2014,Hearin2016}). In this context, it has become an observational  challenge to prove the existence of the so-called {\it{galaxy assembly bias}}, i.e., the dependence of galaxy clustering on secondary halo properties such as the accretion history of halos (see, e.g., \citealt{miyatake2016,dorta2017,niemiec2018,Lin2016}).

In this paper, we provide state-of-the-art measurements of secondary bias for age, spin, and concentration over four orders of magnitude in mass,  in the virial mass range $10.7 \le \log_{10}$ (M$_{vir}$/$h^{-1}$ M$_{\odot}$) $\le 14.7$. This large dynamical range is achieved by combining 4 different MultiDark N-body numerical simulations\footnote{http://skiesanduniverses.org}. In addition, we test for the first time the application of a full multi-tracer approach to the measurement of secondary bias. This technique is based on the fact that different tracers of LSS (e.g., distinct types of halos) occupying the same cosmological volume reflect the same underlying density field. Multi-tracer techniques are designed to minimize the statistical uncertainties associated with cosmic variance by combining the information from distinct biased tracers of the LSS \citep{Seljak:2008, McDonald:2008, Abramo:2013, Abramo:2016}  -- however, our method does not rely on knowledge of the density field, hence it can also be applied to real data.

This paper is organized as follows. Section~\ref{sec:sims} provides a brief description of the MultiDark simulations. The standard method used to measure the relative bias from the correlation function is presented in Section~\ref{sec:methods}. Our main measurements of secondary bias are shown in Section~\ref{sec:secbias}. These measurements are compared with those obtained from a multi-tracer technique in Section~\ref{sec:raul}, where this approach is also briefly described (an extended description of the method can be found in the Appendix). Finally, we compare 
our results with previous literature and summarize the main conclusions of our analysis in Section~\ref{sec:discussion}. Throughout this work, we assume the standard $\Lambda$CDM cosmology \citep{planck2014}, with parameters $h = 0.677$, $\Omega_m = 0.307$, $\Omega_{\Lambda} = 0.693$, $n_s = 0.96$, and $\sigma_8 = 0.823$.

\section{Simulations}
\label{sec:sims}

In order to study the clustering of halos with different secondary properties we use the 
publicly available suite of MultiDark cosmological N-body simulations \citep{multidark2016}. 
In this work we analyze four different simulation boxes: Small MultiDark Planck (SMDPL),
MultiDark Planck 2 (MDPL2), Big MultiDark Planck (BigMDPL) and Huge MultiDark Planck (HugeMDPL).
These boxes have $\sim$4000$^3$ particles and side lengths of 0.4, 1, 2.5 and 4 Gpc/h.
A summary of the numerical parameters of each simulation is shown in Table \ref{table:sims}.

Halos were identified using the ROCKSTAR software \citep{rockstar2013} and we only use the halo catalog at redshift z=0. Furthermore, only distinct halos were included in this analysis. A halo is said to be distinct if its center does not lie within a larger halo. The halo mass function and the velocity function for all distinct halos in each MultiDark simulation are displayed in Figure~\ref{fig:mass_function}. We emphasize in Figure ~\ref{fig:mass_function} the large dynamical range addressed in this analysis. Figure~\ref{fig:mass_function} also illustrates the halo mass and velocity incompleteness of each box (see \citealt{Comparat2017} for further details).   

In this study, we focus on the following halo properties:

\begin{enumerate}
  \item Virial mass, M$_{vir}$, computed in ROCKSTAR using the virial threshold of \cite{Bryan1998}, see  \cite{rockstar2013} for more details.
  \item Maximum circular velocity, V$_{\text{max}}$, defined as
  \begin{equation}
    V^2_{\text{max}} = \text{max} \left[ \frac{G M(<r) }{r} \right]
  \end{equation}
  \item Age, $a_{1/2}$, defined as the scale factor at which half of the peak mass of the halo was accreted.
  \item Spin, $\lambda$, defined as in \cite{bullock2001}, namely:

  \begin{equation}
    \lambda = \frac{|J|}{\sqrt{2} M_{vir} V_{vir} R_{vir}},
  \end{equation}
  where J is the halo's angular momentum and $V_{vir}$ is its circular velocity at the virial radius $R_{vir}$.

  \item Concentration, $c_{200}$, defined as
  
  \begin{equation}
  c_{200} = \frac{R_{200}}{R_{s}},
  \end{equation}
  where $R_{200}$ is computed from
  
  \begin{equation}
  	M_{200} = \frac{4\pi}{3} 200 \rho_{cr} R^3_{200},
  \end{equation}
	and $R_s$ is the Klypin scale radius \citep{klypin2011}, which includes $V_{max}$ and $M_{vir}$ in its definition and assumes a NFW profile \citep{nfw1997}.
\end{enumerate}

\begin{figure}
  \centering
    \includegraphics[width=0.5\textwidth]{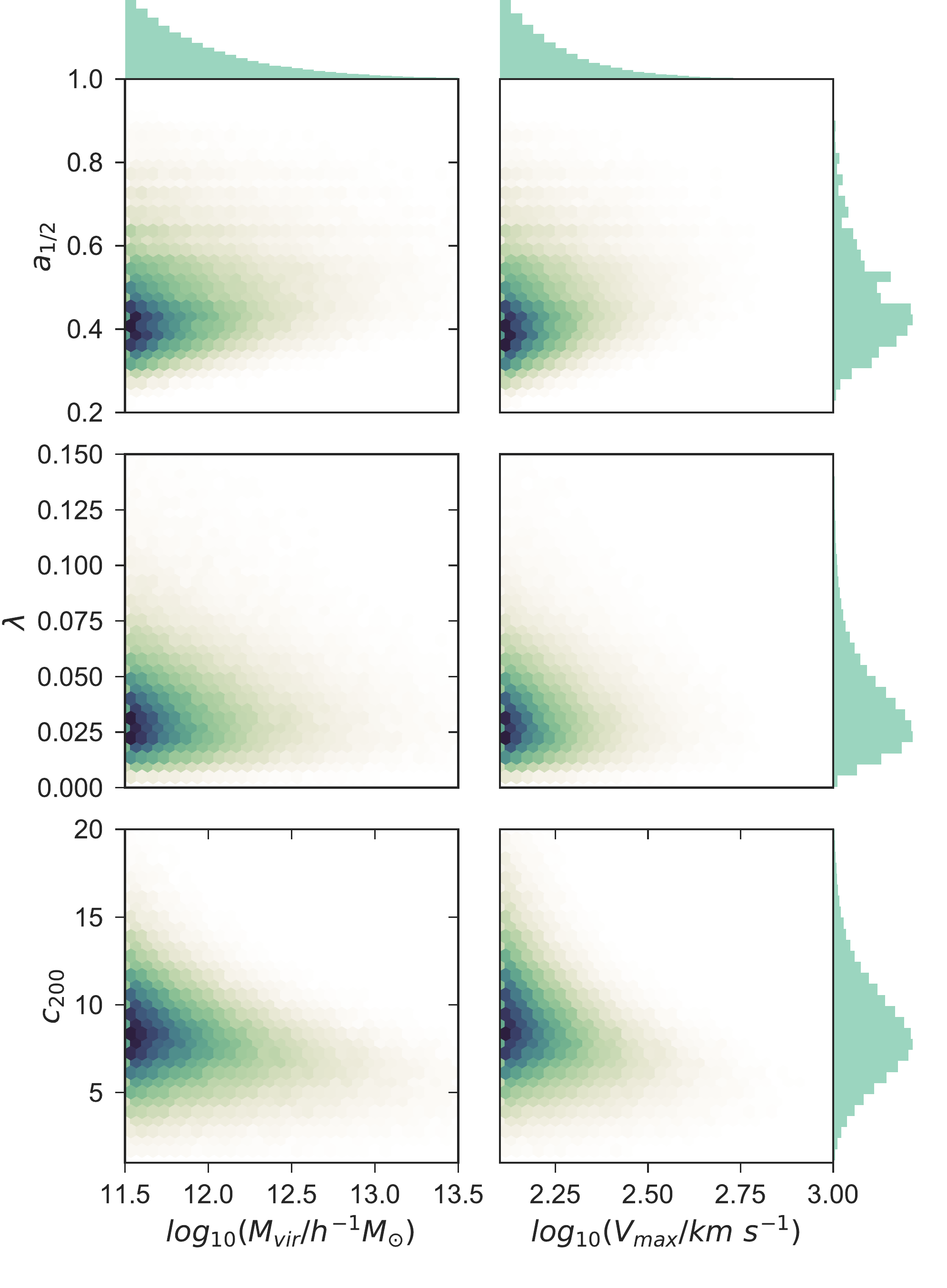}
  \caption{Joint histograms of $M_{vir}$, $V_{max}$, $a_{1/2}$, $\lambda$, and $c_{200}$ and their marginal distributions. Each panel shows the correlation between a primary property and a secondary property in the MDPL2 simulation.}
  \label{fig:prop_corr}
\end{figure}

\begin{table}
   \begin{center}
   \caption{Cuts in virial mass performed in each simulation. These cuts were chosen so that only halos with more than 500 particles are included in the analysis.}
   \begin{tabular}{c|c|c}
      \hline
      Name & Mass cut (h$^{-1}$ M$_{\odot}$) & V$_{max}$ cut (Km s$^{-1}$)\\ \hline
      SMDPL & $ > 4.8 \times 10^{10}$ & $> 80$\\
      MDPL2 & $ > 7.5 \times 10^{11}$ & $> 200$\\
      BigMDPL & $ > 1.2 \times 10^{13}$ & $> 316$ \\
      HugeMDPL &  $ > 4.0 \times 10^{13}$ & $> 631$\\ \hline
   \end{tabular}
   \label{table:mcuts}
  \end{center}
\end{table}

\begin{figure*}
\centering
  \includegraphics[width=\textwidth]{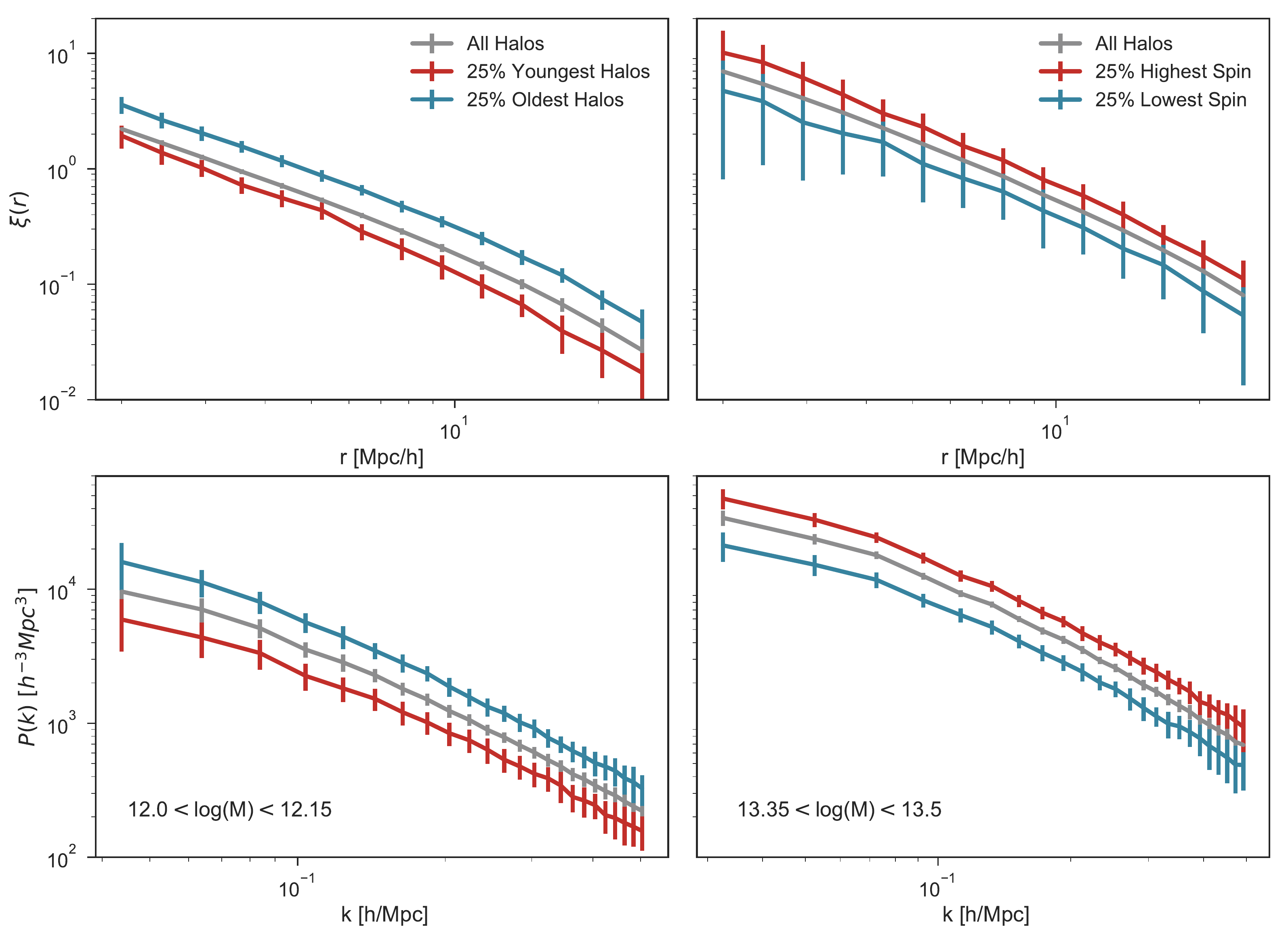}
  \caption{2-point Correlation Function (upper panels) and power spectra (lower panels) illustrating the effect secondary bias. The left panels show the effect of age bias for halos of masses between 10$^{12}$ and 10$^{12.15}$ $h^{-1}$ M$_{\odot}$ in the MDPL2 simulation. The right panels show the effect of spin bias for halos of masses ranging between 10$^{13.35}$ and 10$^{13.5}$ $h^{-1}$ M$_{\odot}$ in the BigMDPL simulation.}
  \label{fig:corrfunc}
\end{figure*}

We consider not only $M_{vir}$ as the primary halo property, i.e. the main predictor of halo clustering, but also $V_{max}$, which characterizes the depth of the gravitational potential well. Note that $V_{max}$ is defined unambiguously in N-body numerical simulations, in contrast to the virial mass, which depends on the particular density-contrast threshold adopted. In addition, adopting $V_{max}$ has been shown to provide important advantages in the context of halo-galaxy connection frameworks such as halo abundance matching \citep{Conroy2006,Trujillo2011}.    
To ensure a robust measurement of all relevant halo properties, we limit our analysis to halos with more than 500 particles. An independent $V_{max}$ cut was implemented, such that less than 1\% of halos in each sample possesses less than 500 particles. For each box, this is equivalent to performing the mass and velocity cuts shown in Table \ref{table:mcuts}.

To illustrate the general features of our dataset, we show in Figure \ref{fig:prop_corr} the distributions of the properties discussed above in the MDPL2 box. The age parameter, 
$a_{1/2}$,  ranges from $\sim$0.3 to $\sim$0.5, which correspond to redshifts of 2.3 and 1,
respectively. Spins are typically in the range $\sim$0.01-0.05, and concentrations span values 
of 5-15, approximately. As expected, very little correlation between secondary and primary 
halo properties is found.

\section{Relative Bias Measurement}
\label{sec:methods}

\begin{figure*}
\centering
    \includegraphics[width=\textwidth]{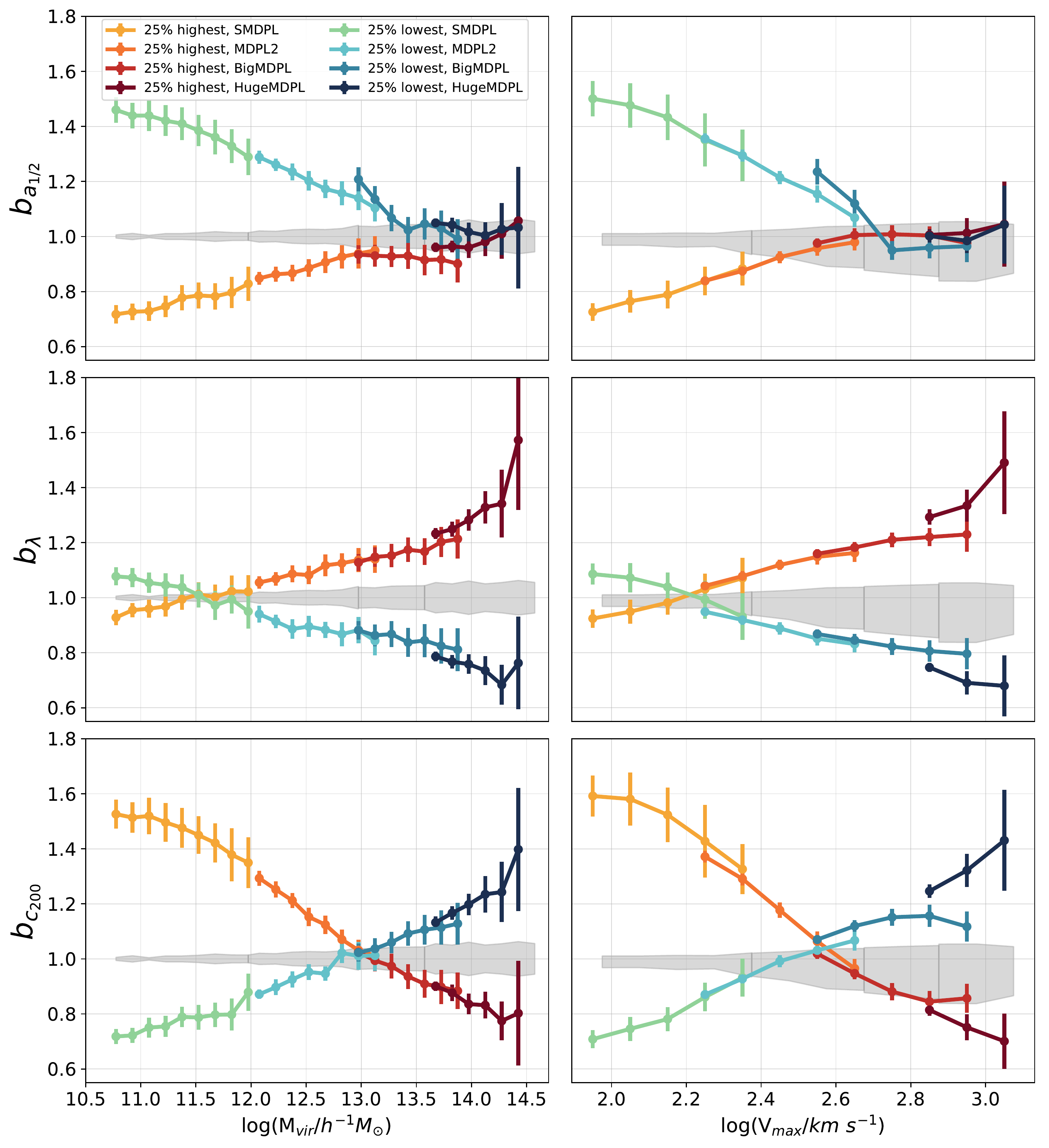}
    \caption{Secondary biases b$(S,P)$ as defined in equation \ref{eq:bias}, estimated using the standard correlation-function method, for primary properties M$_{vir}$ (left column) and V$_{max}$ (right column), and secondary properties a$_{1/2}$, $\lambda$, and c$_{200}$. Each point corresponds to the average among all sub-boxes of a particular simulation and the errors are the standard deviation computed from the entire set of sub-boxes. Note that cross-correlations between primary-property bins have been taken into account in the determination of error bars. All plots show the biases of top and bottom quartiles of a particular secondary property, for all available MultiDark simulations (see text).}
    \label{fig:bias}
\end{figure*}

In this section, we provide a brief description of the standard procedure used to measure secondary bias from simulations, which is based on the computation of the 2-point correlation function. 

To quantify the dependence of halo clustering on a secondary property S, we measure the relative bias between a subsample of halos selected according to S and all halos in the same primary bias property (B) range,

\begin{equation}
   b^2(r, B, S) = \frac{\xi(r, B, S)}{\xi(r, B)},
   \label{eq:bias}
\end{equation}
where we consider the primary bias parameters $B=M_{vir}$ and $V_{max}$, and the secondary parameters $S = a_{1/2}$, $\lambda$, $c_{200}$. 

Each simulation box at $z=0$ was divided in sub-boxes with $L_{sub-box} = L_{box}/4$. The resulting sub-catalogs were further divided in bins of width 0.15 in log$_{10}$(M$_{vir}$) or 0.1 in log$_{10}$(V$_{max}$). The halos with the 25\% highest and lowest values of a particular secondary property were then selected in order to compute the expression shown in equation \ref{eq:bias}.

The 2-point correlation function was measured using CORRFUNC \citep{corrfunc2017}. In the top panels of Figure~\ref{fig:corrfunc}, we show the correlation function for halos selected according to age (left) and spin (right), in the log(M$_{vir}$) ranges 12-12.15, and 13.35-13.5, respectively.   
Here, each point corresponds to the mean value of all sub-boxes and the error bars represent the standard deviation, computed from the entire set of sub-boxes. As found by several previous studies, at fixed halo mass, halos that assemble a significant portion of their mass at earlier times are more tightly clustered than those that assemble at later times. Additionally, halos with a higher spin value are more strongly clustered than low-spin halos. Note that the magnitude of these effects strongly depends on the particular halo mass range selected. In the lower panels of Figure~\ref{fig:corrfunc}, the power spectrum for the same subsets of halos as measured using the multi-tracer approach is also provided (see Section~\ref{sec:raul}). 

To fit the linear bias parameter we include both auto and cross-correlations between different primary property bins, for distances ranging from 5-10 Mpc/h. We choose this range of scales due to the higher signal-to-noise in the assembly bias detection and to facilitate the comparison with previous literature. We then take the ratios of the correlation functions $\xi_{B,S}(r)/\xi_B(r) \rightarrow b^2_{B,S}/b^2_B $ and perform a minimum-$\chi^2$ estimation for the relative bias.

\section{The effect of Secondary bias}
\label{sec:secbias}

The relative biases b$(B,S)$ for all combinations of primary properties $B=M_{vir}, V_{max}$ and secondary properties $S = a_{1/2}, \lambda, c_{200}$ are presented in Figure \ref{fig:bias}. In each plot, the relative bias for halos with the 25\% highest and lowest values of each secondary property, with respect to the entire population in the corresponding primary-property bin, is shown across all available MultiDark simulations. Error bars represent the standard deviation computed from all sub-boxes, where cross-correlations between primary-property bins have been taken into account. Note that the error on the mean is, in most cases, smaller than the size of the markers. 

The shaded region in each plot corresponds to the maximum finite width error, which can potentially be produced by the size of the mass bins. This effect could arise if a secondary property is correlated with halo mass. Thus, when selecting the top and bottom quartiles of such secondary property, one would also be selecting halos based on mass. In order to quantify this effect, we select the 25\% highest and lowest mass values in each mass bin and compute the relative bias. 

As shown in the two upper panels of Figure \ref{fig:bias}, the assembly bias signal at the low-mass end (M $\sim 10^{10.7}$ h$^{-1}$M$_{\odot}$) corresponds to an effect of $\sim 45\%$ for old halos and of $\sim 30\%$ for young halos. The assembly bias detection is consistent with zero at M$\sim 10^{14}$ h$^{-1}$M$_{\odot}$ and beyond, in agreement with recent findings from \cite{mao2018}. The same behavior is seen in the right column, where the primary property is $V_{max}$  
and the effect vanishes at $\log_{10} V_{max}\gtrsim 2.7$.

In the third panel of Figure \ref{fig:bias}, we present results on concentration bias, characterized by the secondary property $c_{200}$. In agreement with previous literature, a significant secondary bias signal is found for c$_{200}$ in the same mass range (see, e.g., \citealt{wechsler2006,gao2007,salcedo2018,han2018}). Here, an inversion occurs at masses $\sim 10^{13}$ h$^{-1}$M$_{\odot}$. At the high-mass end, the difference in bias between the two quartiles is of a factor of 1.75.

Another feature of interest can be seen in the secondary bias signal for spin, $\lambda$, shown in the second row of Figure \ref{fig:bias}. By extending our analysis to very low halo masses, using the SMDPL box, we are able to detect, for the first time, an inversion similar to that found for concentration, with the top and bottom quartiles this time crossing over at masses of $\sim 10^{11.5}$ h$^{-1}$M$_{\odot}$ (or, equivalently, $V_{max} \sim 10^{2.2} km/s$).  The 
large scale of the spin bias effect at the high-mass end, reaching a factor 2 at $\sim 10^{14.5}$ h$^{-1}$M$_{\odot}$, is also noteworthy.

Figure~\ref{fig:bias} confirms previous results regarding the asymmetric nature of secondary bias for age and concentration (see, e.g., \citealt{salcedo2018} for a recent work). The effect appears significantly less pronounced for spin, for which it is only really noticeable at the very high-mass end. A thorough discussion of our results in the context of previous literature is presented in Section~\ref{sec:discussion}.

\section{The multi-tracer approach}
\label{sec:raul}

\begin{figure*}
\centering
    \includegraphics[width=\textwidth]{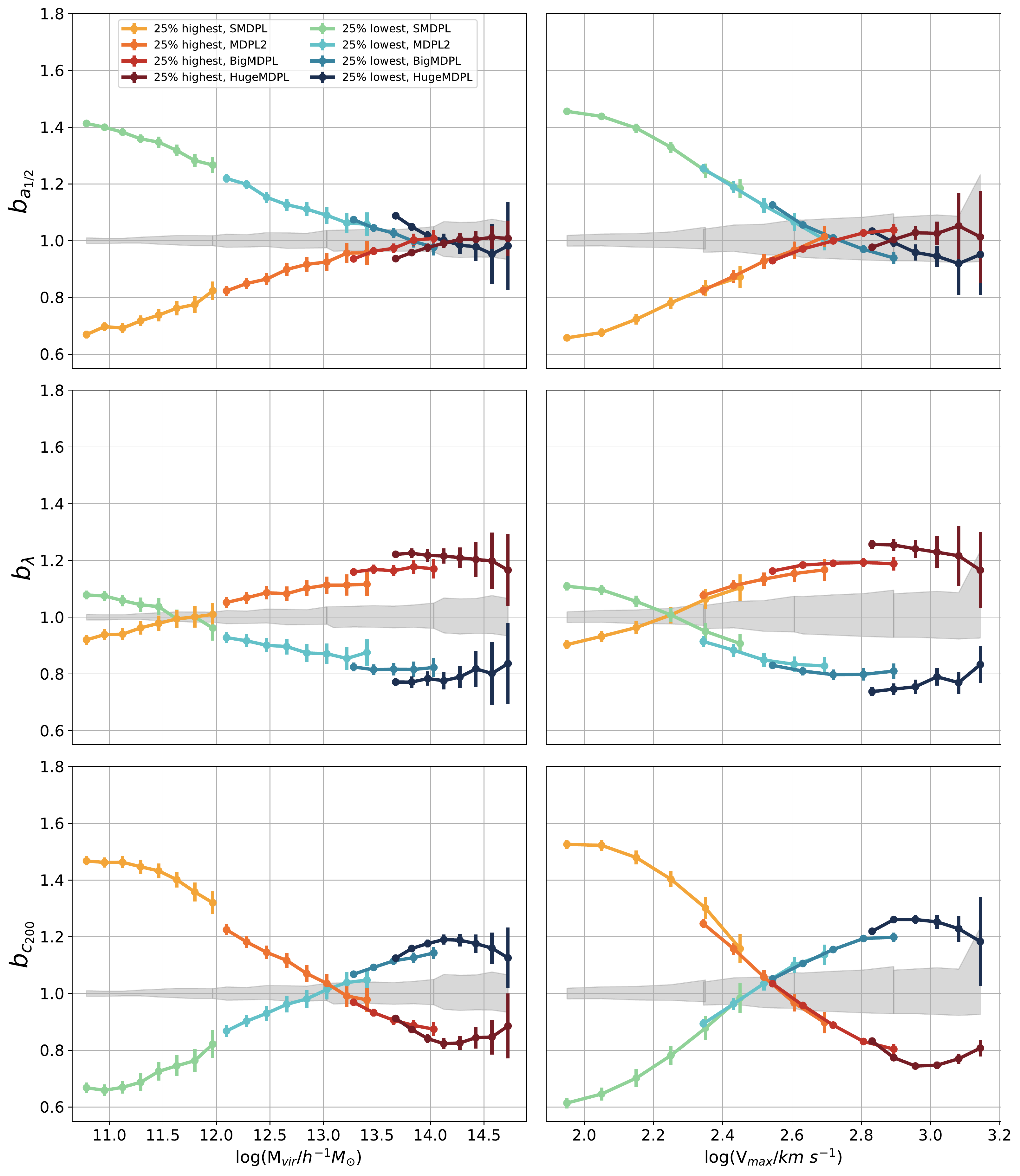}
    \caption{Secondary bias measured through the multi-tracer method, as described in Section \ref{sec:raul}. Similarly to Figure \ref{fig:bias}, the left column corresponds to the primary property M$_{vir}$, the right column to the primary property V$_{max}$, and each row corresponds to the secondary properties a$_{1/2}$, $\lambda$, and c$_{200}$. Once again, each point show the average among all sub-boxes and the errors are the standard deviation from the entire set of sub-boxes.}
    \label{fig:bias_mt}
\end{figure*}

Multi-tracer techniques are designed to minimize the statistical uncertainties associated with cosmic variance by combining the information from distinct biased tracers of the LSS, including halos \citep{Seljak:2008, McDonald:2008, Abramo:2013, Abramo:2016}. They have been employed in the analysis of real data \citep{Blake:2013,Ross2014,Marin2016}, as well as in forecasts for future surveys, where they are expected to be especially beneficial due to the broad nature of their galaxy selections \citep{Ferramacho:2014,2015PhRvD..92f3525A,Fonseca:2015,2017PhRvD..96l3535A,2018arXiv180803093W}. In this work, we have applied multi-tracer techniques to the measurement of secondary bias from the MultiDark simulations, and have regarded each subset of halos, defined through primary and secondary properties, as a different LSS tracer. A brief description of the method can be found in the Appendix (for more information, see \citealt{Abramo:2016}). 

In order to compute the effect of the secondary-bias parameters using power spectra we use the same MultiDark boxes and sub-boxes that were described in Section \ref{sec:methods}, with the same primary bias ($M_{vir}$ or $V_{max}$) bins and same criteria for splitting the samples into secondary-bias classes. We then compute the multi-tracer power spectra for a grand total of 32 different species of tracers -- 8 bins in halo mass, times 4 bins in the secondary bias parameters. 

The linear bias parameter is finally computed in terms of the ratios of spectra $ P_{B,S}(k)/P_B(k) $. The variance of those ratios is determined from the sample of 27 sub-boxes in the case of the 400 $h^{-1}$ Mpc box ($L_{sub-box}=L_{box}/3$), and 64 sub-boxes 
($L_{sub-box}=L_{box}/4$) in the cases of the 1 $h^{-1}$ Gpc, 2.5 $h^{-1}$ Gpc, and 4 $h^{-1}$ Gpc boxes. We fit a constant (linear) bias to the ratios of spectra as a function of bandpower, using all available scales up to $ k \leq {\rm Min} (0.3 \, h \, {\rm Mpc}^{-1},k_{Shot})$ where $k_{Shot}$ is the scale at which the shot noise of the tracer species $(B,S)$ dominates over the power spectrum of that tracer at the level of 90\%, i.e., $P_{B,S} (k_{Shot}) = 0.1 \, P^{Shot}_{B,S} = 0.1 \, \bar{n}_{B,S}^{-1}$. 

In the lower panels of Figure~\ref{fig:corrfunc}, we show the power spectra computed using the multi-tracer method for halos selected according to age (left) and spin (right), in the same halo mass bins used for the correlation function measurements presented in the top panels. As expected, the effects of assembly and spin bias are also significant in the power spectrum.    

Figure \ref{fig:bias_mt} shows the same secondary-bias measurements of Figure \ref{fig:bias}, but determined through the multi-tracer method. We immediately notice that the multi-tracer method yields significantly smaller statistical errors across all MultiDark simulations when measuring secondary biases. Our derived uncertainties in the secondary bias range from $\lesssim 1\%$ for the least massive (and most abundant) halos of the SMDPL box, to $\sim 5 - 10\%$ for the massive halos of the HugeMDPL box. When compared to the standard method, we find that the decrease in uncertainty depends on secondary property, quartile, and mass. For the MDPL2 simulation, as a reference, we report a decrease of roughly 25-40\% in the uncertainty for age as the secondary property. For spin the decrease is typically around 10-35\% and for concentration we find a decrease of up to 30\%.

We must emphasize that the results presented in Figs. \ref{fig:bias} and \ref{fig:bias_mt} are not directly comparable, since they probe different distance scales. In particular, the bias estimated through the correlation function is measured within a narrow distance range of $5 - 10$ $h^{-1}$ Mpc, whereas the bias in terms of the power spectrum is measured on scales $k \sim 0.05 - 0.3 \, h \, {\rm Mpc}^{-1}$, depending on the box size and level of shot noise. In particular, this means that, if there is a significant scale dependence of the secondary bias, the measurements using correlation function and power spectra could differ. In that case our results using the power spectra may differ from tracer to tracer, and from box to box -- on that regard, see the Appendix. 

With the aforementioned caveats, both methods provide qualitatively similar results for assembly, concentration and spin bias. Some small differences are, however, worth noting. In particular, the multi-tracer measurements appear flatter at the high-mass end, especially for HugeMDPL, as compared to the standard measurements. In addition, the top-right panel of Figure~\ref{fig:bias_mt} seems to show hints of a cross-over between young and old halos for high-V$_{max}$ halos. Further investigation will be required in order to clarify the origin of these differences.

\section{Discussion \& Conclusions}
\label{sec:discussion}

By combining all available MultiDark simulations, we provide the most precise measurement of secondary halo bias across a wide virial mass range, spanning from $10^{10.7}$ to $10^{14.7} h^{-1} M_{\odot}$. Furthermore, we present the first multi-tracer measurement of secondary bias across the same comprehensive dynamical range.

For masses below $10^{14} h^{-1} M_{\odot}$ we find, in agreement with, e.g., \citet{gao2005}, \citet{wechsler2006}, \citet{li2008}, \citet{salcedo2018}, \citet{mao2018} and \citet{chue2018}, that older halos are more strongly clustered than younger halos, for the definition of age given in Section \ref{sec:sims}. Above $10^{14} h^{-1} M_{\odot}$ we detect no solid evidence of halo assembly bias or of a cross-over between the top and bottom age quartiles, in agreement with \citet{mao2018}, who also used the MDPL2 box. Interestingly, some hints of a cross-over are found when V$_{max}$ is used as the primary halo property and the simulation data is analyzed using the multi-tracer technique. This is not observed when a traditional approach is employed, nor when halo mass is used as the primary halo property in the multi-tracer measurement. Follow-up work will be devoted to clarify the origin of this feature in V$_{max}$. We highlight that, using the BigMDPL simulation, \citet{chue2018} reports a significant assembly bias signal at $10^{15} h^{-1} M_{\odot}$. Through the methodology adopted in this paper, however, we were unable to obtain statistically significant measurements in the same mass range. 

We present a novel feature in the secondary bias produced by spin: a cross-over between top and bottom quartiles at $M \sim 10^{11.5} h^{-1} M_{\odot}$ or, equivalently, at $V_{max} \sim 10^{2.2} km/s$. We are able to probe such small halo masses due to the high mass resolution of the SMDPL simulation. Although the presence of a cross-over in age is still uncertain, the detection of this phenomenon in spin, in combination with previous detections for concentration and several other properties, might be an indication that the inversion of the secondary bias signal is a universal property. In this context, \citet{salcedo2018} argues that the behavior of spin bias is different from that of other secondary biases, which may suggest a distinct physical origin. Some of the evidence we present here, however, might cast serious doubt on such claim. For example, \citet{salcedo2018} states that spin bias differs from other secondary biases due to its weak mass dependence and the fact that it increases with mass, whereas most other secondary biases decrease with mass. However, in the wider mass range probed in this work, we find that the relative bias for spin varies as much as for age ($\sim 0.5$). Moreover, the detection of a cross-over places spin bias in the same footing as other secondary biases, since it shows that the fact that it increases with mass was an artifact of the mass scales that were being probed.

Our results place updated constraints to models addressing the connection between galaxies and halos. The wide dynamical range that we probe encompasses  halos hosting from emission-line galaxies at the low-mass end to massive quiescent galaxies and clusters at the high-mass end. In the context of the halo-galaxy connection, it has become an observational  challenge to prove the existence of {\it{galaxy assembly bias}}, a hypothesis that states that the clustering properties of galaxies, at fixed halo mass, depend on secondary halo properties such as the accretion history (or age) of halos (see, e.g.,\citealt{miyatake2016,dorta2017,niemiec2018,Lin2016}). For luminous red galaxies (LRGs), indications of the existence of galaxy assembly bias have been reported by \cite{dorta2017} and \cite{niemiec2018}, from a sample extracted from the Baryon Oscillation Spectroscopic Survey (BOSS, \citealt{Dawson2013}). The typical mass of the halos inhabited by these galaxies is estimated in $\sim 10^{12.7-13}$ h$^{-1}$M$_{\odot}$ (\citealt{niemiec2018}), but a large scatter is expected, so that the most massive LRGs could inhabit halos above $\sim 10^{14}$ h$^{-1}$M$_{\odot}$. These results are, in principle, compatible with both the halo assembly bias and the concentration bias signal reported in Figure~\ref{fig:bias}. Our measurement of halo assembly bias at the high-mass end can be properly connected to the results presented in \cite{dorta2017} and \cite{niemiec2018} through the technique of ``age matching" \citep{Hearin2016}. 

In addition, the large scale of the spin bias signal for cluster-size halos, of almost a factor 2 at $\sim 10^{14.5}$ h$^{-1}$M$_{\odot}$, suggests an alternative route towards an observational proof of secondary bias. Probing the effect of spin bias requires measuring rotation, or a proxy for it, in a large sample of galaxy clusters. These measurements,  which are still extremely challenging, might be possible in the near future thanks to techniques such as kinetic Sunyaev-Zel'dovich effect (see, e.g., \citealt{Baldi2018}).   

We have applied, for the first time, a fully multi-tracer approach to the measurement of secondary bias. Here, different subsets of halos are viewed as different LSS tracers. We show that this method is capable of reproducing all features observed from the standard measurement. In addition, we have obtained statistical uncertainties which are comparable to other bias estimates from N-body simulations that minimize cosmic variance by relying on direct knowledge of the density field, e.g. \citet{gao2007,chue2018}. However, in contrast to those methods, which in effect compute $\delta_\alpha/\delta_m$ at each point (in configuration or Fourier space), the technique used in this work can be applied both to simulations as well as to actual data, in real or redshift space.

The multi-tracer approach is a promising technique for future surveys like Euclid{\footnote{https://www.euclid-ec.org}}, or DESI{\footnote{https://www.desi.lbl.gov}}, where cosmological measurements will be performed from data sets containing multiple galaxy populations. The effort presented is this paper will serve as the basis for future developments aimed at the application of the method to real data.    

The main conclusions of this work can be summarized as follows: 

\begin{itemize}
	\item No statistically significant halo assembly bias signal (secondary bias on $a_{1/2}$) is detected for halos above $M \sim 10^{14} h^{-1} M_{\odot}$. 
    
	\item A cross-over is detected, for the first time, in the spin bias signal. Below $M \sim 10^{11.5} h^{-1} M_{\odot}$, lower-spin halos are more tightly cluster than higher-spin halos, at fixed halo mass. The signal reverses above this characteristic mass. The effect of spin bias increases significantly at the high-mass end, reaching almost a factor of 2 at $M \sim 10^{14.5} h^{-1} M_{\odot}$.
	
	\item We test, for the first time, the performance of a fully multi-tracer approach for the measurement of secondary bias. These techniques are designed to minimize the statistical uncertainties associated with cosmic variance by combining information from distinct biased tracers of the LSS, in this case, halos.   
    
    \item The multi-tracer approach is capable of reproducing all secondary-bias features observed from the standard measurement, with the advantage that the signal-to-noise improves significantly. We find that the decrease in uncertainty depends on secondary property, quartile, and mass. For the MDPL2 simulation, as a reference, we report a decrease of roughly 25-40\% in the uncertainty for age as the secondary property. For spin the decrease is typically around 10-35\% and for concentration we find a decrease of up to 30\%.
    
	\item Our halo assembly bias measurements are consistent with results suggesting that galaxy assembly bias can be detected from massive galaxies alone, and, in particular, from LRGs.     	
\end{itemize}

\section*{Acknowledgments}

GSP and ADMD thank FAPESP for financial support. LRA thank both FAPESP and CNPq for financial support. 
FP acknowledges support from the Spanish MINECO grant AYA2010-21231-C02-01.

We thank New Mexico State University (USA) and Instituto de Astrof\'isica de Andaluc\'ia CSIC (Spain) for hosting the Skies \& Universes site for cosmological simulation products.

We also thank Peter Behroozi for running Rockstar Consistent Tree on the MultiDark simulations.

\bibliographystyle{mnras}
\bibliography{references}

\appendix
\section{The multi-tracer technique}

\begin{figure*}
\centering
    \includegraphics[width=\textwidth]{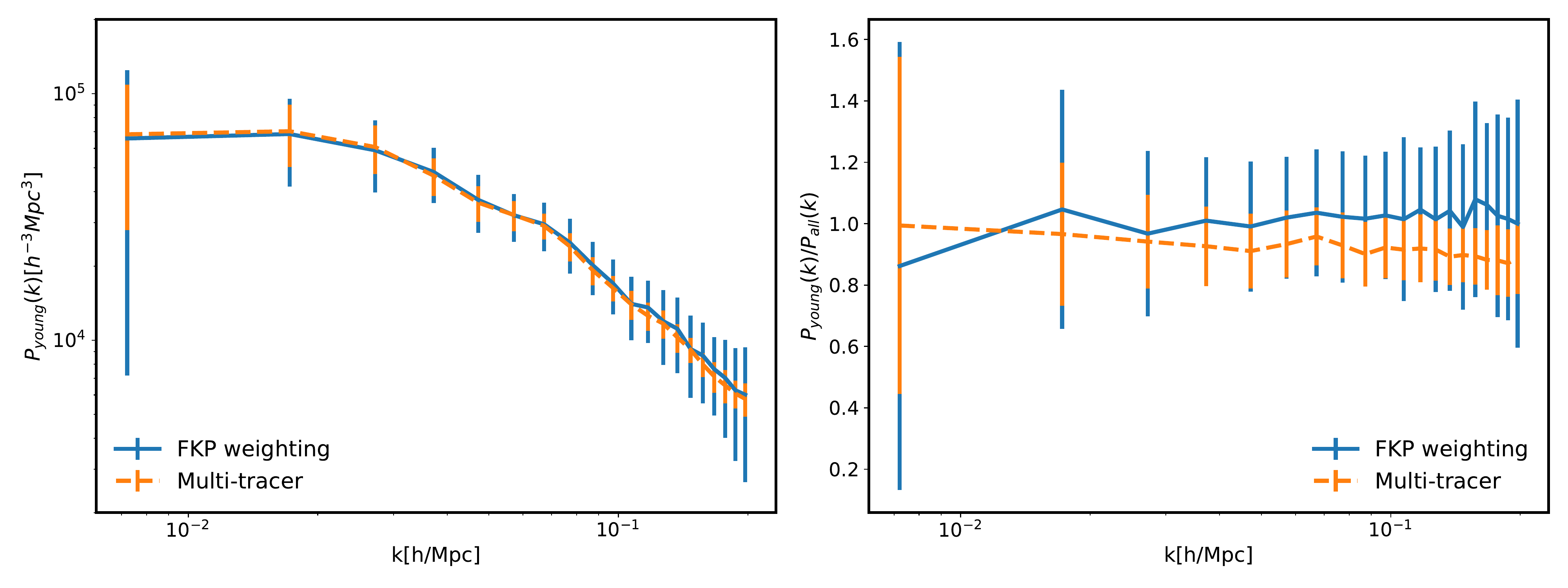}
    \caption{Left: Power spectrum in the entire wavenumber range considered for a subset containing the youngest halos in the virial mass bin $\log_{10} M [h^{-1} \, M_\odot] = 13.825 \pm 0.075$, extracted from the 4-Gpc HMDPL box. Dashed lines correspond to the power spectra computed using the multi-tracer optimal weighting, whereas solid lines show the traditional (FKP) weighting. Right: Same as before for the ratio of the power spectrum of the youngest halos and that of the entire population in the corresponding bin.}
    \label{fig:appendix1}
\end{figure*}

The optimal combination of the density field of any number of biased tracers, which results in minimum-variance estimators for the power spectra of those tracers, was derived in \citet{Abramo:2016}. The technique consists in a generalization of the weights obtained by \citet{FKP} for a single tracer -- the so-called FKP weighting. The multi-tracer weighted density field of a tracer species $\alpha$ is defined as $f_\alpha (\vec{x}) = \sum_\beta w_{\alpha \beta} (\vec{x}) \delta_\beta (\vec{x})$, with the weights:
\begin{equation}
\label{eq:mtw}
w_{\alpha \beta} = \bar{n}_\alpha b_\alpha \, \delta_{\alpha \beta}  - \frac{\bar{n}_\alpha b_\alpha^2 P_m }{1+{\cal{P}}_{eff}} \bar{n}_\beta b_\beta \; ,
\end{equation}
where $\bar{n}_\alpha (\vec{x})$ is the measured number density of the tracer $\alpha$, $b_\alpha$ is its fiducial bias, $P_m=P_m(k_p)$ is the fiducial matter power spectrum at the pivot scale $k_p=0.1 h$ Mpc$^{-1}$, and ${\cal{P}}_{eff} = \sum_i \bar{n}_i b_i^2 P_m$.
We then compute the Fourier transform of the weighted fields, take their (quadratic) amplitudes, combine them into bandpowers, and transform these quadratic forms into estimator for auto-power spectra of the individual tracers -- see \citet{Abramo:2016} for details.

The weighted fields are used to compute quadratic estimators, and those are then combined into the multi-tracer power spectra. Finally, we take the ratios of the spectra of each one of those 32 tracers to the multi-tracer spectra of their respective mass bins, $P_{B,S}(k)/P_{B}(k) \to b_{B,S}^2 (k)/b_B^2 (k)$. 

A key aspect of multi-tracer optimal weighting is the fact that the weighted fields are linear combinations of the density fields of each tracer, hence the multi-tracer power spectrum estimators automatically include all the auto- and cross-spectra of all tracers, in a minimum-variance combination. In particular, this means that the information from cross-spectra has already been taken into account by the multi-tracer auto-spectra estimators, and we do not need to compute them separately.

In Figure~\ref{fig:appendix1}, we show the power spectrum (left panel) for a particular age quartile (youngest halos) in the mass bin $\log_{10} M [h^{-1} \, M_\odot] = 13.825 \pm 0.075$, extracted from the 4-Gpc HMDPL box, as well as the ratio of this power spectra to the power spectrum of the entire mass bin (right panel). The dashed lines correspond to the power spectra computed using the multi-tracer optimal weighting, whereas the solid lines show the traditional (FKP) weighting. It is clear that the variances of the amplitudes of the power spectra, as well as the statistical fluctuations of the ratios of spectra (the relative bias), are smaller when we employ the multi-tracer weighting. In the case of this mass bin and box, the spectra are dominated by shot noise, so the multi-tracer technique, which uses the information from the cross-correlations, is naturally a less noisy estimator of the power spectrum. When shot noise is less relevant (as happens for low-mass bins), the statistical fluctuations of the power spectra estimated with the two methods can be similar, but the covariance of the ratios of the power spectra are typically much smaller when we employ the multi-tracer method \citep{Abramo:2016}.

\label{lastpage}

\end{document}